# Pressure-Induced Superconducting-like Transition in the *d*-wave Altermagnet Candidate CsV₂Se₂O

Yuanzhe Li[1,2†], Yilin Han[1,2†], Liu Yang[1,2†], Wanli He[1,2†], Pengda Ye[1,2†],

Wencheng Huang[3,4†], Jiabin Qiao[1,2,*],

Yuemei Li[1,2], Xiaodong Sun[1,2], Tingli He[1,2], Jiayi Han[1,2], Yuxiang Chen[1,2],

Ruifeng Tian[1,2], Hao Sun[1,2], Yuwei Liu[1,2], Feng Wu[1,2], Baoshan Song[1,2],

Zhengtai Liu[5], Mao Ye[5], Yaobo Huang[5], Kenichi Ozawa[6], Ji Dai[7],

Massimo Tallarida[7], Shengtao Cui[8], Jie Chen[1,2], Meiling Jin[1,2], Wayne Zheng[1,2],

Chaoyu Chen[3,4,*], Zhiwei Wang[1,2,*], Zhi-Ming Yu[1,2,*], Xiang Li[1,2,*], Yugui Yao[1,2,*]

[1]*Centre for Quantum Physics, Key Laboratory of Advanced Optoelectronic Quantum Architecture and Measurement (MOE), School of Physics, Beijing Institute of Technology, Beijing, 100081, China.*

[2]*Beijing Key Laboratory of Quantum Matter State Control and Ultra-Precision Measurement Technology, School of Physics, Beijing Institute of Technology, Beijing, 100081, China.*

[3]*Songshan Lake Materials Laboratory, Dongguan 523808, China*

[4]*Department of Physics, Southern University of Science and Technology, Shenzhen 518055, China*

[5]*Shanghai Synchrotron Radiation Facility, Shanghai Advanced Research Institute, Chinese Academy of Sciences, Shanghai, China*

[6]*Institute of Materials Structure Science, High Energy Accelerator Research Organization (KEK), Tsukuba, Ibaraki 305-0801, Japan*

[7]*ALBA Synchrotron, Carrer de la Llum 2-26 08290 Cerdanyola del Vallès, Barcelona, Spain*

[8]*National Synchrotron Radiation Laboratory, University of Science and Technology of China, Hefei, China*

† These authors contributed equally to this work.

* Corresponding authors.

Emails:

jiabinqiao@bit.edu.cn

chenchaoyu@sslab.org.cn

zhiweiwang@bit.edu.cn

zhiming_yu@bit.edu.cn

xiangli@bit.edu.cn

ygyao@bit.edu.cn

**Abstract**

**Altermagnetism generates exchange-type spin splitting without net magnetization and, in its $d$-wave form, resembles the angular symmetry of unconventional $d$-wave superconductivity. Whether this correspondence bears directly on superconducting instabilities in real correlated materials remains open. Here we study the quasi-two-dimensional vanadium oxychalcogenide $CsV_2Se_2O$ (CVSO), a square-net $d$-wave altermagnet candidate, through combined experimental and theoretical investigation of its lattice structure, electronic structure and transport properties. At ambient pressure, CVSO is a weakly insulating parent state with a density-wave-like anomaly near 100 K, and its bulk properties are most consistent with a G-type compensated antiferromagnetic background. Under compression, the density-wave-like feature is suppressed, the magnetoresistance evolves from predominantly negative to positive, and a superconducting-like resistive downturn emerges below about 3 K. This low-temperature anomaly is reproducible across samples and pressure media, and is suppressed by magnetic field. Room-temperature X-ray diffraction reveals no symmetry lowering, whereas does show a pronounced compressibility anomaly over the same pressure range. CVSO thus reveals a pressure-tuned phase diagram in which a reconstructed weakly insulating parent state gives way to strange-metal-like transport and superconducting-like behavior, echoing broader phenomenology associated with unconventional superconductors, including cuprates and nickelates.**

Altermagnetism has recently emerged as a distinct form of long-range magnetic order beyond the conventional ferromagnetic–antiferromagnetic dichotomy. Its defining characteristic is momentum-dependent spin splitting in a magnetically compensated state, enforced by spin group symmetry even in the non-relativistic limit[1-8]. In its $d$-wave form[9-12], the reciprocal-space spin texture shares the same angular symmetry as the order parameter of unconventional $d$-wave superconductors[11-26]. This correspondence extends the significance of altermagnetism beyond magnetic classification alone to a broader issue in correlated-electron physics: how momentum-selective spin splitting in compensated antiferromagnets reshapes the low-energy electronic landscape in which superconductivity may eventually emerge.

Among the candidate materials, the alkali vanadium oxychalcogenides $AV_2Ch_2O$ (A = K, Rb, Cs; Ch = S, Se, Te) stand out[11,12,26-39]. Their $V_2O$ square-net layers adopt an anti-$CuO_2$ motif, providing a structural analogue of the cuprate planes while embedding the V sublattice in a symmetry environment conductive to $d$-wave altermagnetism. Members of this family, including $KV_2Se_2O$ (KVSO), $RbV_2Te_2O$ (RVTO), $CsV_2Te_2O$ (CVTO), and $CsV_2Se_2O$ (CVSO), span a broad range of correlated ground states in which compensated antiferromagnetism, low-energy reconstruction and electronic correlations meet within a common square-net framework[11,12,35-38]. Modest changes in chemistry, stoichiometry and interlayer coupling are therefore sufficient to reorganize the low-energy state.

KVSO illustrates both the promise of this family and the present uncertainty about its microscopic ground state. Early spectroscopic and theoretical work interpreted KVSO in terms of a room-temperature $d$-wave altermagnetism and associated the anomaly near 105 K with a density-wave reconstruction on top of a so-called C-type antiferromagnetic configuration, that is, antiferromagnetic within the $V_2O$ plane but ferromagnetically stacked along $c$ axis[11]. Subsequent neutron diffraction instead favored bulk G-type antiferromagnetism, with nearest-neighbor spins antiparallel along all three crystallographic directions, and suggested additional low-temperature modulation[31]. At the same time, the low-energy transport of KVSO is highly sensitive to stoichiometry and sample condition, with reports spanning weakly metallic and weakly insulating behavior[29,34,37]. Consistent with this sensitivity, atomic-scale measurements on near-stoichiometric crystals reveal checkerboard antiferromagnetic order together with a long-range $\sqrt{2} \times \sqrt{2}$ lattice modulation and a low-temperature gap at the Fermi level $(E_F)$[37]. Across the broader vanadium-oxychalcogenide family, related anomalies have also been connected to the vanadium electronic reconfiguration[32] or to Lifshitz-like Fermi-surface changes[35]. These results point to a close coupling between compensated antiferromagnetism and low-energy electronic reconstruction, while leaving the microscopic character of the reconstructed state under active debate.

CVSO enters this landscape from a more robustly reconstructed and non-metallic side of the phase space. Earlier work on polycrystalline CVSO reported weakly insulating transport together with a density-wave-like anomaly already at ambient pressure[26]. More recent scanning tunneling microscopy/spectroscopy (STM/STS) measurements have further implicated $d$-wave altermagnetic order in its low-energy state[38]. CVSO therefore accesses the same intertwined ingredients, namely square-net magnetism, low-energy reconstruction and strong correlation, but from a more reconstructed parent state than the more sample-dependent KVSO case. This contrast becomes especially informative in light of recent high-pressure work on KVSO, where compression suppresses metallic behavior and instead drives a metal–insulator transition without superconductivity[34]. Most recently, ambient-pressure superconductivity has been reported in a related non-centrosymmetric structural variant, $Na_{2-x}V_2Se_2O$, with a maximum reported transition temperature of 16.3 K[40], broadening the superconducting landscape of $V_2Se_2O$-based layered oxychalcogenides while still leaving open how a reconstructed parent state in stoichiometric $AV_2Ch_2O$ evolves under a clean tuning parameter such as pressure.

Here we combine scanning transmission electron microscopy (STEM), angle-resolved photoemission spectroscopy (ARPES), transport and magnetization measurements, high-pressure X-ray diffraction (XRD), and density functional theory (DFT) calculations to study CVSO across ambient and high pressure. We show that ambient-pressure CVSO is a weakly insulating parent state with a density-wave-like anomaly near 100 K, whose bulk properties are most consistent with a G-type compensated antiferromagnetic background. Under compression, the density-wave-like feature is suppressed, the normal state is reorganized, and a low-temperature superconducting-like response appears. The resulting pressure-tuned phase diagram, in which a reconstructed weakly insulating parent state gives way to strange-metal-like transport and superconducting-like behavior, places this square-net altermagnet candidate in close phenomenological dialogue with broader unconventional-superconductivity phase diagrams.

**Ambient-pressure parent state of CVSO**

We begin by establishing the crystal and bulk physical properties of CVSO at ambient pressure. CVSO crystallizes in the tetragonal layered structure with space group $P4/mmm$ (Fig. 1a). Its basic structural unit is the anti-CuO$_2$-type V$_2$O square net, in which Se atoms reside above and below the centers of the V$_2$O plaquettes and Cs layers separate adjacent V$_2$Se$_2$O slabs. Atomic-resolution high-angle annular dark-field scanning transmission electron microscopy (HAADF-STEM) images taken along two orthogonal zone axes resolve the expected stacking sequence and atomic arrangement directly (Fig. 1b, c). The projected atomic columns and the extracted lattice parameters, $a = b = 4.00$ Å and $c = 7.90$ Å, are consistent with the tetragonal structure extracted from our X-ray diffraction (Extended Data Fig. 1) and previous reports[27,39]. No impurity phase is observed even in large-field-of-view STEM images, attesting to the high crystalline quality of the crystals.

The bulk transport simultaneously shows that CVSO, unlike more sample-dependent members such as KVSO[29,34,37], lies robustly on the non-metallic side. The longitudinal resistivity $\rho_{xx}(T)$ increases continuously on cooling and exhibits a clear anomaly at $T_{DW}$ near 100 K, more clearly resolved in its temperature derivative (Fig. 1d). Importantly, this behavior is observed here in single crystals and closely matches earlier measurements on polycrystalline CVSO[26], indicating that the non-metallic normal state is intrinsic rather than an artefact of sample granularity. At low temperature, $\rho_{xx}(T)$ follows an approximately logarithmic temperature dependence over the measured range. Conventional thermally activated[41], variable-range-hopping[42] and small-polaron-hopping descriptions[43], however, do not account for the data, and the field dependence likewise disfavors weak-localization[44] and Kondo scenarios[45,46] (see Methods and Extended Data Figs. 2, 3). The low-temperature transport is therefore better understood as a weakly insulating regime associated with density-wave-like reconstruction. The magnetic susceptibility ($\chi$) exhibits a weak anomaly on the same temperature scale, but remains small throughout the measured temperature range (Fig. 1e). No separate low-temperature magnetic transition is resolved, indicating that the feature near 100 K develops within a pre-existing antiferromagnetic background rather than marking the onset of long-range antiferromagnetism itself.

To determine the magnetic background underlying this parent phase, we considered the two collinear antiferromagnetic configurations most relevant to this family: the C-type state, in which spins are antiparallel within the $V_2O$ plane but ferromagnetically stacked along the $c$ axis, and the G-type state, in which nearest-neighbor spins are antiparallel along all three crystallographic directions. Our total-energy calculations show that the G-type configuration is energetically favored over the C-type one (Fig. 1h, upper; Extended Data Fig. 4), consistent with recent neutron scattering results of KVSO and CVTO[31,35,36]. At the level of a single $V_2O$ layer, this G-type background is compatible with a hidden altermagnetic description of the low-energy states[36], while the bulk crystal remains a compensated antiferromagnet. The corresponding G-type calculation yields several bands crossing the $E_F$ and multiple Fermi-surface contours in the Brillouin zone (Fig. 1f, 1i and Extended Data Fig. 4). These calculated Fermi-surface contours find direct counterparts in the ARPES Fermi-surface intensity map (Fig. 1g), and the main measured dispersions are likewise captured by the G-type calculation.

To describe the additional low-temperature reconstruction, we further include a density-wave-like order in the calculation (Fig. 1h, lower), namely a translational-symmetry-breaking reconstruction that folds the original bands into a reconstructed Brillouin zone and opens partial hybridization gaps near $E_F$[11]. In particular, along the Γ–X, X–M and M–Γ segments, the reconstructed calculation reveals avoided crossings and partial near-Fermi-surface hybridization gaps. These momentum-selective folded features and gap openings are directly resolved in ARPES spectra taken at 20 K, that is, below $T_{DW}$ (Fig. 1j). The partial gap opening observed at selected momenta may reduce the low-energy spectral weight available for charge transport, and thus further supports the speculation that the density-wave-like reconstruction contributes to the weakly insulating behavior at low temperature. Therefore, the calculation-experiment correspondence supports an ambient-pressure state in which bulk CVSO is governed by a G-type compensated antiferromagnetic background and undergoes an additional low-temperature density-wave-like reconstruction.

**Iso-structural phase transition in pressurized CVSO**

To track the structural evolution of CVSO under compression, we performed high-pressure X-ray diffraction (XRD) measurements at room temperature using a diamond anvil cell (Fig. 2a). The pressure-dependent diffraction map is shown in Fig. 2b. Across the full pressure range investigated, all diffraction features shift systematically towards higher $2\theta$ with increasing pressure, as expected from lattice compression, whereas no new reflections or obvious peak splitting emerge. Representative refinements at low, intermediate and high pressures are shown in Fig. 2c (see also Extended Data Fig. 5). In all cases, the diffraction profiles are indexed consistently within the tetragonal $P4/mmm$ structure, indicating that the average crystallographic symmetry is preserved throughout compression.

The extracted lattice parameters nevertheless reveal anomalous compressional behavior. As shown in Fig. 2d, both $a$ and $c$ decrease continuously with pressure, but neither follows a single linear trend over the full pressure range. The $c$ axis is substantially more compressible at low pressure and stiffens above the shaded interval $P_c$, whereas the $a$ axis evolves more gradually. As a result, the axial ratio $c/a$ exhibits a clear kink over the same pressure window. The unit-cell volume $V$ likewise cannot be captured by a single smooth equation of state. Instead, separate third-order Birch–Murnaghan (BM) fits describe the low- and high-pressure regimes, with a clear deviation in the intermediate region (Methods). This behavior points to a pronounced redistribution of compressibility without an accompanying change in average crystallographic symmetry.

To examine this feature more directly, we analyzed the reduced pressure $H$ as a function of Eulerian strain $f_E$ within the standard linearized BM formalism[47]. For uniform compression within a single structural regime, $H(f_E)$ is expected to follow an approximately linear relation. As shown in Fig. 2e-g, however, the reduced pressures derived from the $a$ axis, the $c$ axis and the unit-cell volume all deviate from a single linear trend and instead display a pronounced anomaly within 15 ~ 22 GPa. The deviation is strongest in the axial- and volume-derived quantities, consistent with the anomalous redistribution of axial compressibility already evident in Fig. 2d. Together with the absence of new diffraction peaks, peak splitting or detectable symmetry lowering, these results support possibly an iso-structural transition occurring over this pressure interval[47-49]. This pressure window provides the structural backdrop for the transport evolution discussed below.

**Suppression of the density-wave-like instability under pressure**

We next examine the compression-driven evolution of the ambient-pressure density-wave-like instability. Figure 3a shows the temperature dependence of the longitudinal resistivity $\rho_{xx}$ for sample S1 measured under a set of pressures (see also Extended Data Fig. 6 for more data for other samples). At low pressure, CVSO retains the weakly insulating behavior observed at ambient pressure, with resistivity increasing upon cooling. A clear anomaly remains visible, consistent with the density-wave-like transition identified in the ambient-pressure state (Fig. 1d). With increasing pressure, the overall resistivity is progressively reduced, the low-temperature upturn becomes weaker, and anomaly shifts to lower temperature and becomes less distinct. As shown in Fig. 3b, c, the characteristic temperatures $T_{DW}$ extracted from normalized differential resistivity $\rho^*$ for S1 and S2 both decrease monotonically with pressure and become difficult to resolve above a critical pressure range. The continuous evolution of $\rho_{xx}$ and its derivative therefore suggests a gradual suppression of the density-wave-like state.

Further insight comes from the pressure evolution of the magnetoresistance (MR) for S1 and S2, as shown in Fig. 3d and 3e, respectively. At low pressure, the MR is predominantly negative over the measured field range. With increasing pressure, the magnitude of the negative MR is progressively reduced and eventually changes sign, such that the high-pressure response becomes positive and more characteristic of itinerant carriers. This crossover points to a substantial redistribution of the underlying transport channels as the density-wave-reconstructed state is weakened.

To quantify this evolution, we fitted the field dependence of the MR using a phenomenological model containing a positive quadratic term associated with the conventional orbital response (term I) and two negative terms (terms II and III) describing field-suppressed scattering channels dominant in the density-wave regime[50-54] (Methods). The model reproduces the data well across the full pressure range investigated (Extended Data Fig. 7). The normalized weights of the three fitting components are summarized in Fig. 3f. At low pressure, transport is dominated by the two negative contributions (terms II and III), whereas with increasing pressure these are rapidly suppressed and the positive orbital term (term I) becomes dominant. Notably, this redistribution occurs mainly within 15 ~ 22 GPa, closely overlapping with the pressure interval extracted from the structural analysis (Fig. 2e-g) and with the pressure range over which the density-wave-like anomaly becomes strongly weakened and eventually difficult to resolve (Fig. 3b, c). The suppression of the reconstructed state, the reorganization of transport channels and the iso-structural anomaly therefore unfold over a common pressure window.

**Emergence of superconducting-like transitions under high pressure**

Once the density-wave-like feature is sufficiently weakened, a new low-temperature anomaly appears in transport. As shown in Fig. 4a, a resistive downturn emerges under pressure from about 10 GPa and remains observable over a broad pressure interval. This feature is absent in the lower-pressure regime where the density-wave-like anomaly is still pronounced, indicating that it develops only after substantial weakening of the competing reconstructed state.

The low-temperature resistive drop is strongly field sensitive. As shown in Fig. 4b, the anomaly shifts systematically to lower temperature with increasing magnetic field and is gradually suppressed, whereas the normal-state background is much less affected. Comparable behavior is observed in different samples and under different pressure-transmitting media (Fig. 4c), indicating that the effect is reproducible and not primarily controlled by pressure conditions. These features are consistent with a superconducting-like response. To further quantify its field dependence, we extracted the transition temperature at each magnetic field and analyzed the resulting $\mu_0 H_{c2}(T)$ using the three-dimensional Ginzburg-Landau framework[55] (Fig. 4d). The fitted upper critical fields remain finite throughout the pressure range over which the low-temperature anomaly is observed, further supporting the presence of a pressure-induced superconducting-like regime.

The evolution of the characteristic temperatures is summarized in the temperature–pressure phase diagram in Fig. 4e. The density-wave-like anomaly decreases monotonically with pressure, whereas the superconducting-like transition appears only after the former has been strongly weakened and persists over a finite pressure interval. In the intervening region, the normal state becomes metallic and its resistivity is well described by a power-law form with an exponent close to unity (Extended Data Fig. 8), pointing to a strange-metal-like regime rather than a conventional Fermi-liquid metal[56]. The phase diagram therefore indicates that the superconducting-like response develops as the density-wave-reconstructed parent state collapses and gives way to a strange-metal-like phase under pressure.

## Discussion

The ambient-pressure phenomenology of CVSO is most consistently described in terms of a correlated, density-wave-reconstructed weakly insulating parent state. The low-temperature upturn is unlikely to arise from a simple semiconducting gap or standard disorder localization alone. Rather, it reflects a correlated background that is further reconstructed by the density-wave-like instability near 100 K, which suppresses low-energy states available for charge transport and thereby enhances the weakly insulating character of the normal state. We therefore identify ambient-pressure CVSO as a correlation-dominated weakly insulating state embedded in a compensated antiferromagnetic background.

A microscopic picture consistent with both DFT and experiment starts from the V $d_{xy}$, $d_{yz}$ and $d_{xz}$ orbitals and a corresponding multi-orbital description of the two-sublattice V–O square net. Within this framework, cooling below $T_{\mathrm{DW}}$ introduces an itinerant density-wave instability that selectively gaps nested portions of the Fermi surface and folds the band structure, naturally accounting for the partial gap opening and folded ARPES features. In this sense, the weakly insulating transport, the density-wave-like anomaly, and the reconstructed low-energy electronic structure are all aspects of the same correlated parent state.

Under pressure, this reconstructed state is progressively weakened, as reflected by the common pressure scale of the compressibility anomaly, the suppression of $T_{\mathrm{DW}}$, and the MR-channel redistribution. Because the room-temperature diffraction patterns remain indexed within the same average $P4/mmm$ framework under compression, the main role of pressure is better understood as a clean tuning of bandwidth, hybridization and interaction balance within a fixed structural manifold, rather than as the driver of a conventional symmetry-lowering structural transition. The superconducting-like downturn therefore emerges not from a simple good metal, but only after the reconstructed or weakly insulating parent state has been substantially weakened.

In this respect, CVSO exhibits a phase-diagram evolution analogous at the phenomenological level to that of cuprates[57] and nickelates[58]: a reconstructed parent state is first suppressed, a strange-metal-like regime intervenes, and superconducting-like transport appears only on the high-pressure side of this broader electronic reorganization. We emphasize, however, that this similarity does not by itself imply an identical microscopic pairing mechanism.

An elemental-impurity origin, such as superconductivity in elemental V, Se or Cs, can also be ruled out[59-62]. The superconducting-like response in CVSO is therefore more naturally viewed as developing intrinsically from a still reconstructed and spatially inhomogeneous electronic background. This places CVSO near a broader class of pressure-tuned correlated systems in which pairing tendencies appear before the competing ordered background is completely removed. In that regime, residual spin fluctuations associated with the compensated antiferromagnetic background and/or density-wave-related correlations provide a plausible source of pairing interactions, although the pairing symmetry, the degree of bulk coherence, and the possible coexistence with residual density-wave order all remain open questions.

In summary, CVSO reveals a pressure-tunable route from a density-wave-reconstructed weakly insulating parent state, through a strange-metal-like regime, to a superconducting-like state. The present results therefore place this square-net system in close phenomenological dialogue with broader unconventional-superconductivity phase diagrams, while leaving open the key microscopic questions of pairing symmetry, bulk coherence, and possible coexistence with residual density-wave order.

## Methods

### Sample synthesis

$CsV_2Se_2O$ single crystals were grown by a flux method with CsSe as the flux. A starting mixture of Cs, Se, V, and $V_2O_5$ was prepared with a molar ratio of $CsV_2Se_2O$ : CsSe = 1 : 5. This mixture was loaded into an alumina crucible, and the crucible was sealed inside an evacuated quartz tube. The sealed tube was first heated to 40 °C and kept 1 hour for pre-reaction. Afterwards, the temperature was heated to 1000 °C and held for 10 hours to ensure a homogeneous melt. The system was then cooled slowly to 650 °C at a rate of 2 °C/h. Excess flux was finally removed by centrifugation, yielding pure $CsV_2Se_2O$ single crystals.

### Scanning transmission electron microscopy

Atomic-resolution high-angle annular dark-field (HAADF) images were acquired using a spherical-aberration-corrected JEOL ARM200F scanning transmission electron microscope (STEM) equipped with double aberration correctors and operated at 200 kV. The TEM specimen oriented along the [100] zone axis was prepared by focused ion beam (FIB) milling, whereas the specimen along the [00$\bar{1}$] zone axis was prepared by mechanical exfoliation using Scotch tape.

### Single-crystal X-ray diffraction

Single-crystal X-ray diffraction data were collected on a BRUKER D8 VENTURE diffractometer equipped with the APEX 5 software and Mo radiation. A specimen with approximate dimensions $0.1 \times 0.04 \times 0.02$ mm$^3$ was isolated and protected under an N$_2$ atmosphere during data collection.

### Powder X-ray diffraction

Powder X-ray diffraction (XRD) data were collected on a BRUKER D2 PAHSER diffractometer using Cu Kα radiation under ambient conditions. Air-sensitive samples were loaded into an air-sensitive sample rack inside a glove box prior to measurement. Crystal structure Rietveld refinements were performed using the Fullprof package.

**Angle-resolved photoemission spectroscopy**

ARPES measurements were performed at the BL03U beamline of the Shanghai Synchrotron Radiation Facility (SSRF) equipped with a Scienta Omicron DA30 energy analyzer and p-polarized radiation. The samples were cleaved in situ under base pressure better than $9 \times 10^{-11}$ mbar and temperature below 15 K. During the experiment, the beam spot size was set to $15 \times 15$ μm$^2$. The energy resolution was set to 10 ~ 20 meV depending on the photon energy used, and the angular resolution was set to $0.2°$. Supporting ARPES measurements are also performed at the BL09U of SSRF, BL28A of photon factory, BL20 of ALBA synchrotron, and BL13U of National Synchrotron Radiation Laboratory, Hefei.

**High-pressure X-ray diffraction and equation-of-state analysis**

High-pressure powder X-ray diffraction measurements were performed at room temperature in a diamond anvil cell. CsV$_2$Se$_2$O powder sample was ground into a fine powder inside a glove box, then pressed using an empty DAC. The pressed sample was cut into $0.15 \times 0.15 \times 0.01$ mm$^3$ pieces and loaded into an optical DAC. The high-pressure X-ray diffraction data were collected on a BRUKER D8 VENTURE diffractometer equipped with the APEX 5 software and Ag radiation. Diffraction patterns were collected over the full pressure range investigated, and Rietveld refinements were performed in the Fullprof package within the tetragonal *P4/mmm* structure to extract the lattice parameters and unit-cell volume.

The pressure dependence of the unit-cell volume $V$ was fitted using the third-order Birch-Murnaghan equation of state (BM EoS) expressed as[63]:

$$P(V) = \frac{3B_0}{2} \left[ \left( \frac{V_0}{V} \right)^{\frac{7}{3}} - \left( \frac{V_0}{V} \right)^{\frac{5}{3}} \right] \times \left\{ 1 + \frac{3}{4}(B_0' - 4) \left[ \left( \frac{V_0}{V} \right)^{\frac{2}{3}} - 1 \right] \right\}, \tag{1}$$

where $V_0$ is the zero-pressure unit-cell volume, $B_0$ the bulk modulus, and $B_0'$ its pressure derivative. Separate fits were performed for the low- and high-pressure regimes, excluding the intermediate anomalous interval.

To evaluate the compressibility anomaly, we analyzed the reduced pressure $H$ as a function of the Eulerian strain $f_E$. For the volume analysis[47-49],

$$f_E = \left[(V_0/V)^{\frac{2}{3}} - 1\right]\Big/2,$$ (2)

and

$$H = P\Big/\left[3f_E(1 + 2f_E)^{\frac{5}{2}}\right].$$ (3)

Within the standard linearized Birch-Murnaghan (BM) formalism, the equation of state becomes:

$$H = B_0 + \frac{3}{2}B_0(B_o' - 4)f_E.$$ (4)

An analogous axial analysis was carried out for the $a$ and $c$ lattice parameters, with

$$f_E^{(a)} = \frac{1}{2}\left[\left(\frac{a_0}{a}\right)^2 - 1\right], \; f_E^{(c)} = \frac{1}{2}\left[\left(\frac{c_0}{c}\right)^2 - 1\right],$$ (5)

and the corresponding reduced pressures $H_a$ and $H_c$ were evaluated using the same definition as in Eq. (3). The characteristic pressure interval $P_c$ was identified from the common anomalous region in the $c/a$ ratio, the unit-cell volume evolution and the $H\text{-}f_E$ analyses.

**Transport measurements**

Ambient-pressure (magneto-)transport measurements were primarily carried out in two closed-cycle cryogenic systems (Oxford Instruments TeslatronPT, 1.6–300 K) equipped with superconducting magnets of 12 T and 14 T. Electrical resistance was measured in a standard four-terminal configuration using either an AC lock-in technique (samples S1–S4, excitation current 1 mA at 13.777 Hz) or a DC current source and nanovoltmeter combination (Keithley 6221 and 2182A, delta mode, samples S1–S2, excitation current 1 mA).

High pressure was generated by a diamond anvil cell made of nonmagnetic Be-Cu alloy. Diamond anvils with a 400-μm culet were used. Single crystals with a dimension of $80 \times 80 \times 5 \; \mu m^3$ were loaded together with distinct pressure-transmitting media (NaCl for S1, S2, S4-S6 and silicon oil for S3) into the sample chamber. Pressure was calibrated by the ruby fluorescence method. The temperature dependence of the resistivity was measured under selected pressures down to low temperature. Magnetoresistance (MR) measurements were performed with magnetic field applied perpendicular to the sample plane. The MR is defined as,

$$\text{MR}(B) = [\rho_{xx}(B) - \rho_{xx}(0)]/\rho_{xx}(0). \tag{6}$$

The characteristic temperature $T_{\text{DW}}$ of the density-wave-like anomaly was determined from the minimum in $d\rho_{xx}/dT$. The onset temperature of the superconducting-like transition, $T_c$, was defined as the temperature point at which low-temperature resistive downturn occurs.

**Origin of the low-temperature weakly insulating behavior**

To clarify the origin of the low-temperature upturn in $\rho_{xx}(T)$, we compared the data with several standard transport forms. Over the low-temperature range, the resistivity is approximately described by a logarithmic temperature dependence[26],

$$\rho_{xx}(T) \propto \log(1/T). \tag{7}$$

For comparison, we also fitted the data using activated transport, variable-range hopping, and small-polaron hopping[41-43]:

$$\rho(T) = \rho_0 \exp\left(\frac{E_a}{k_B T}\right), \tag{8}$$

$$\rho(T) = \rho_0 \exp\left[\left(\frac{T_0}{T}\right)^{\frac{1}{4}}\right], \tag{9}$$

$$\rho(T) = \rho_0 \, T \exp\left(\frac{E_a}{k_B T}\right). \tag{10}$$

Here $E_a$ is an activation energy, $T_0$ is the characteristic hopping temperature, and $\rho_0$ is a pre-factor. Although the logarithmic form provides a better empirical description of the low-temperature data, none of the conventional semiconductor- or hopping-based models captures the overall behavior (Extended Data Fig. 2).

We further considered weak-localization and Kondo scenarios, both of which are expected to show characteristic magnetic-field responses. The measured magnetotransport does not support either mechanism in the low-temperature regime (Extended Data Fig. 3). These comparisons disfavor conventional semiconductor, hopping and disorder-driven scenarios, and instead support a weakly insulating regime associated with density-wave-like reconstruction.

**Magnetoresistance analysis**

The field dependence of the magnetoresistance (MR) was analyzed using a phenomenological model,

$$\text{MR}(H) = aH^2 - b\ln(1 + cH^2) - d|H|, \tag{11}$$

where the positive quadratic term $aH^2$ (term I) describes the conventional orbital MR expected for an itinerant metallic response[53,54], whereas the latter two negative terms (terms II and III) capture field-suppressed scattering channels that are most prominent in the SDW-like regime[50-52]. The MR curve at each pressure was fitted independently. To compare the relative evolution of the three transport channels, the fitted amplitudes were normalized by the sum of the absolute values of all three fitted amplitudes at the same pressure (Fig. 3f). These normalized weights were then used to track the crossover from negative-scattering-dominated transport to orbital-response-dominated transport.

**Analysis of the superconducting-like transition and normal-state power law**

To quantify the field dependence of the superconducting-like downturn, the transition temperature extracted at each magnetic field was analyzed within a three-dimensional Ginzburg–Landau framework[55],

$$\mu_0 H_{c2}(T) = \mu_0 H_{c2}(0) \frac{(1 - t^2)}{(1 + t^2)}, \tag{12}$$

where $t = T/T_c$ is the reduced temperature. This analysis was applied to the datasets shown in Fig. 4d. In the present work, the resulting $H_{c2}(0)$ values are used as phenomenological field scales associated with the superconducting-like response.

To characterize the intermediate-pressure normal state, the resistivity above the superconducting-like downturn was fitted to a power-law form[56],

$$\rho_{xx}(T) = \rho_0 + AT^n, \tag{13}$$

where $\rho_0$, $A$ and $n$ are fitting parameters. Values of $n$ close to unity were taken as indicative of strange-metal-like transport, while remaining cautious that the accessible temperature window is limited and the system remains electronically reconstructed.

**Density functional theory calculations**

The calculations were performed in the framework of density functional theory as implemented in the Vienna ab initio simulation package[64]. The projector augmented wave (PAW) pseudopotentials[65] were adopted in the calculation and generalized gradient approximation (GGA)[66] of the Perdew-Burke-Ernzerhof (PBE)[67] functional was selected as the exchange-correlation potential. The Brillouin zone (BZ) was sampled using Monkhorst-Pack k-mesh[68] with a size of 15×15×5. The energy cutoff was set to 600 eV and the energy convergence criteria was $10^{-6}$ eV. We construct a Wannier tight-binding model Hamiltonian with V $d$ orbitals, O $p$ orbitals, and Se $p$ orbitals, by the maximally localized Wannier functions (MLWF) method, and the Fermi surfaces were also calculated via the WANNIER90[69,70].

**Low-energy theoretical framework**

To connect the DFT results with the phenomenological discussion in the main text, the low-energy sector of CVSO may be organized in terms of the V $d_{xy}$, $d_{yz}$ and $d_{xz}$ orbitals on the two-sublattice V–O square net. At the level of a generic interacting model, this sector can be written as:

$$H = \sum_{(ij),\alpha,\beta,\sigma} \left( t_{\alpha\beta} c_{i\alpha\sigma}^{\dagger} c_{j\beta\sigma} + h.c. \right) + \sum_{\alpha} \left( e_{\alpha} \sum_{i} n_{i\alpha} \right) + U \sum_{i,\alpha} n_{i\alpha\uparrow} n_{i\alpha\downarrow}$$

$$+V \sum_{i,\alpha \neq \beta} n_{i\alpha} n_{i\beta} + \cdots, \tag{14}$$

where $\alpha, \beta$ label the orbitals, $U$ is the intro-orbital Hubbard interaction, and $V$ is the inter-orbital interaction.

For the high-temperature unreconstructed electronic structure, a minimal four-band tight-binding Hamiltonian can be constructed by fitting to the DFT-derived bands,

$$H_{AM} = \sum_{i \in A} \left( t_1 c_{i\alpha}^\dagger c_{i+\hat{x},\alpha} + t_2 c_{i\beta}^\dagger c_{i+\hat{x},\beta} + t_3 c_{i\alpha}^\dagger c_{i+\hat{y},\alpha} + t_4 c_{i\beta}^\dagger c_{i+\hat{y},\beta} + h.c. \right)$$
$$+ \sum_{i \in B} \left( t_1 c_{i\gamma}^\dagger c_{i+\hat{y},\gamma} + t_2 c_{i\delta}^\dagger d_{i+\hat{y},\delta} + t_3 c_{i\gamma}^\dagger c_{i+\hat{x},\gamma} + t_4 c_{i\delta}^\dagger c_{i+\hat{x},\delta} + h.c. \right) \quad , \quad (15)$$
$$+ e_1 \left( \sum_{i \in A} c_{i\alpha}^\dagger c_{i\alpha} + \sum_{i \in B} c_{i\gamma}^\dagger c_{i\gamma} \right) + e_2 \left( \sum_{i \in A} c_{i\beta}^\dagger c_{i\beta} + \sum_{i \in B} c_{i\delta}^\dagger c_{i\delta} \right)$$

where $\alpha, \beta, \gamma, \delta = d_{xy}^\uparrow, d_{yz}^\uparrow, d_{xy}^\downarrow, d_{xz}^\downarrow$ denote the basis spin-orbitals used in the model. This low-energy effective Hamiltonian is introduced as a phenomenological starting point for discussing the reconstructed state inferred from experiment.

Within this framework, the anomaly at $T_{\text{DW}}$ is attributed to a density-wave-like instability that reconstructs the low-energy electronic structure, folds the bands, and suppresses spectral weight near the Fermi level. This picture is consistent with the folded ARPES features and partial hybridization gaps discussed in the main text, while leaving open the precise character and commensurability of the underlying order. Both AM above $T_{\text{DW}}$ and the density-wave-like order below $T_{\text{DW}}$ are spontaneously symmetry broken orders from the Hamiltonian $H$.

Pressure modifies hopping, bandwidth, and orbital hybridization, thereby tuning the balance between kinetic energy and interaction-driven reconstruction. Compression can thus weaken the reconstructed state and promote a more itinerant normal-state response, consistent with experiment. Within this picture, superconducting-like transport appears only after the reconstructed parent state has been substantially weakened, possibly in the presence of residual antiferromagnetic and/or density-wave-related fluctuations. This superconducting-like state emerges from the instability as another U(1) spontaneous symmetry breaking, where pairing is mediated by the AM/density-wave-like spin fluctuations. The pairing symmetry, the degree of bulk coherence, and the possible coexistence with residual density-wave-like order still remain elusive, which require further experimental and theoretical investigations.

**Data availability**

Source data are provided with this paper.

**Acknowledgements**

We acknowledge the discussions with Fan Yang and Jin-Guang Cheng. This work was supported by the National Key Research and Development Program of China (Grant Nos. 2025YFA1411200, 2022YFA1403400 and 2022YFA1403700), the Beijing National Laboratory for Condensed Matter Physics (Grant No. 2023BNLCMPKF007), the National Natural Science Foundation of China (NSFC) (Grant Nos. 12574068, 12504176) and Guangdong Basic and Applied Basic Research Foundation (Grant Nos. 2026B0303000004, 2022B1515020046, 2022B1515130005 and 2021B1515130007). J. Q. acknowledge the support by the Beijing Institute of Technology Research Fund Program for Young Scholars.

**Author contributions**

J.Q., Z.W., Z.Y., X.L. and Y. Y. conceptualized the project. L.Y. grew the single crystals. Y.L., P.Y., Y.L., X.S. and B.S. prepared the electric and optical DAC. P.Y., Y.C. and J.H. conducted the X-ray diffraction measurements. Y.L. conducted the transport measurements. Y.H., T.H., Z.Y. conducted the first-principles calculations. W.Z. provided theoretical support. W.H. and C.C contributed to the ARPES measurement and analysis with the support from Z.L, M.Y., Y.H., K.O., J.D., M.T. and S.C.. J.Q., C.C., Z.Y., W.Z., Y.L., Y.H., W.H. and P.Y. analyzed the data and wrote the paper with key input from all authors.

**Competing interests**

The authors declare no competing interests.

insulator and spin singlet state in antiferromagnetic $KV_2Se_2O$. arXiv: 2511.06712.

## Figures

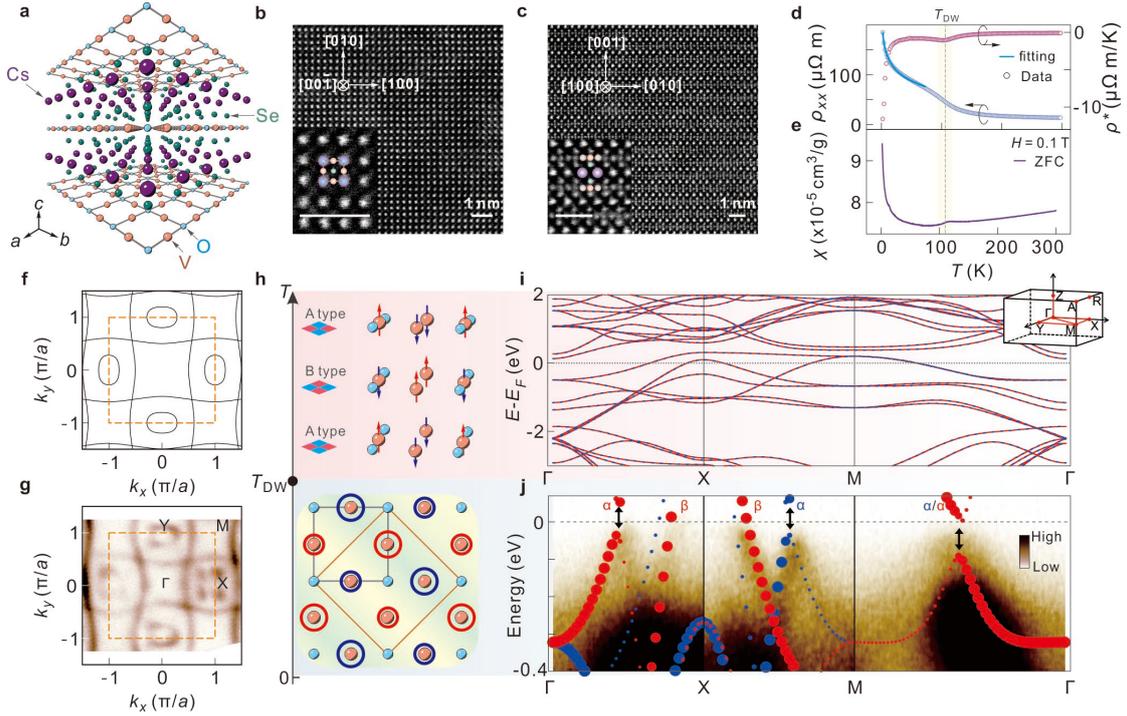

**Figure 1 | Ambient-pressure parent state of CsV₂Se₂O (CVSO). a**, Crystal structure of CVSO at ambient pressure. **b, c**, Atomic-resolution HAADF-STEM images taken along the $[00\bar{1}]$ and $[100]$ zone axes, respectively, confirming the high crystalline quality of the single crystals. **d**, Temperature dependence of the longitudinal resistivity $\rho_{xx}$ and its differential $\rho^* = d\rho_{xx}/dT$, showing a weakly insulating normal state and a clear anomaly at $T_{DW} \sim 100$ K. **e**, Temperature dependence of magnetic susceptibility $\chi$ under 0.1 T with a magnetic field parallel to the $c$ axis. **f**, Calculated Fermi surfaces (FSs) of CVSO in the G-type configuration in the $k_x$-$k_y$ plane. **g**, ARPES Fermi-surface intensity plot at $E_F$ measured at 20 K. The dashed rectangles in **f** and **g** mark the first Brillouin zone. **h**, Schematic illustrations of the magnetic/electronic configurations used in the analysis: the candidate collinear antiferromagnetic states considered for the bulk background (upper) and the low-temperature (below $T_{DW}$) density-wave-like reconstructed state adopted in the calculation (lower). **i**, Calculated band structure for bulk CVSO in the G-type antiferromagnetic state. Red and blue curves denote the spin-up and spin-down bands. **j**, Comparison between the ARPES spectra measured at 20 K and the calculated spin-resolved band structure along $\Gamma - X - M - \Gamma$ for a V₂O layer by taking into account the low-temperature density-wave-like reconstruction (**h**, lower). The red and blue dots denote the spin-up and -down bands, respectively, and their size scales the spectra

weight from the calculations. The black arrows mark the partial hybridization gaps opening near $E_F$ along the $\Gamma$-X, X-M and M-$\Gamma$ segments.

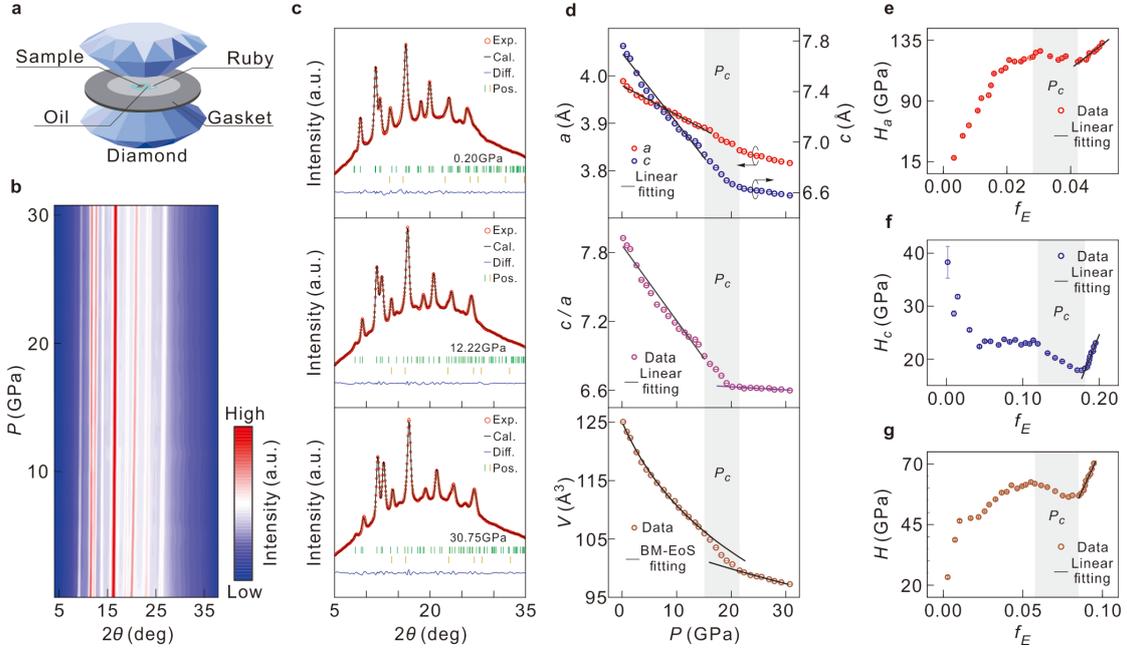

**Figure 2 | Structure evolution of CVSO under pressure. a,** Schematic of the diamond anvil cell used for high-pressure optical measurement. **b,** Powder XRD patterns of CVSO at room temperature and various pressures up to 30.75 GPa. The red and blue colors represent the high and low diffraction intensity, respectively. **c,** LeBail refinement of the XRD profiles collected at 0.20 GPa, 12.22 GPa and 30.75 GPa. Red circles, black and blue curves denote the experimental data (Exp.), calculated results (Cal.) and their difference (Diff.), respectively. Green and golden vertical lines mark the Braggs peaks of $CsV_2Se_2O$ and Au (calibrate pressure). **d,** Pressure dependences of the lattice parameters $a$ and $c$, the lattice-parameter ratio $\frac{c}{a}$, and the unit-cell volume $V$ of $CsV_2Se_2O$. Clear anomalies appear over 15–22 GPa, as marked by the shaded region, consistent with an isostructural compression anomaly. **e–g,** Eulerian-strain analysis of the compression behavior of CVSO. The normalized pressures $H_a$, $H_c$, and $H$ are plotted as functions of Eulerian strain $f_E$. The shaded region marks the same pressure window as in **d**.

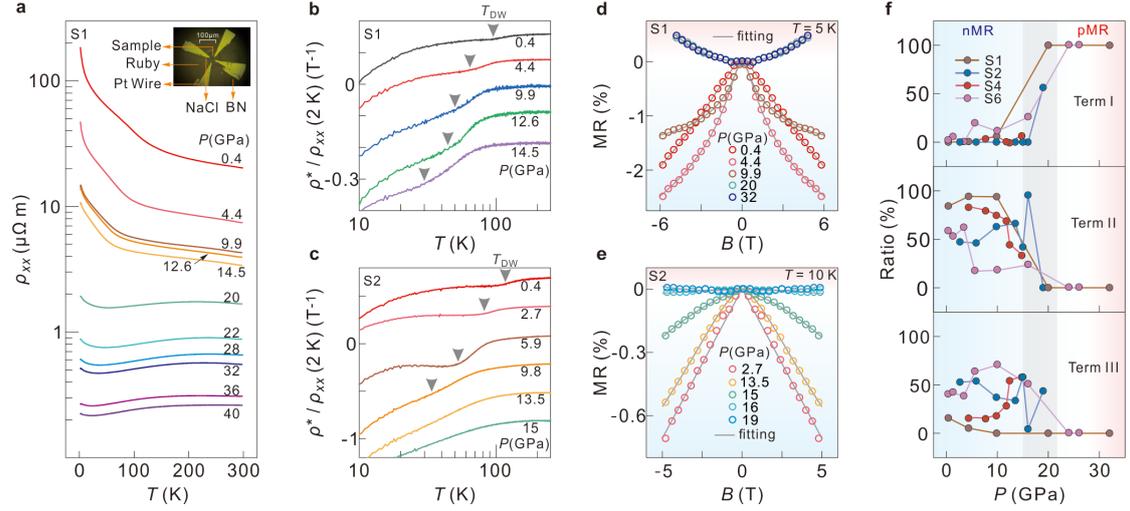

**Figure 3 | Pressure evolution of the density-wave-like state and magnetotransport. a,** Longitudinal resistivity $\rho_{xx}(T)$ from 0.4 to 40 GPa in Sample S1. Inset is the photograph of the diamond anvil cell configuration. **b,** Normalized differential resistivity $\frac{\rho^*}{\rho_{xx}(2\,K)}$ from 0.4 to 14.5 GPa in Sample S1. The gray arrow marks the density-wave-like anomaly temperature $T_{DW}$. **c,** $\frac{\rho^*}{\rho_{xx}(2\,K)}$ from 0.4 to 15 GPa in Sample S2. **d, e,** Pressure-dependent magnetoresistance of S1 and S2 at fixed temperatures. Open circles denote the measured data and the solid grey lines show the corresponding fits. **d.** Magnetoresistance curves of S1 collected at 5 K under different pressures. **e.** Magnetoresistance curves of S2 collected at 10 K under different pressures. Colored empty circles denote the experimental data, and the gray curves are the fits using the phenomenological model or eq. (5). **f,** Relative contributions of the fitted positive orbital term and the two negative field-suppressed terms as a function of pressure for samples S1, S2, S4 and S6. The shaded region highlights the common pressure interval within 15 ~ 22 GPa.

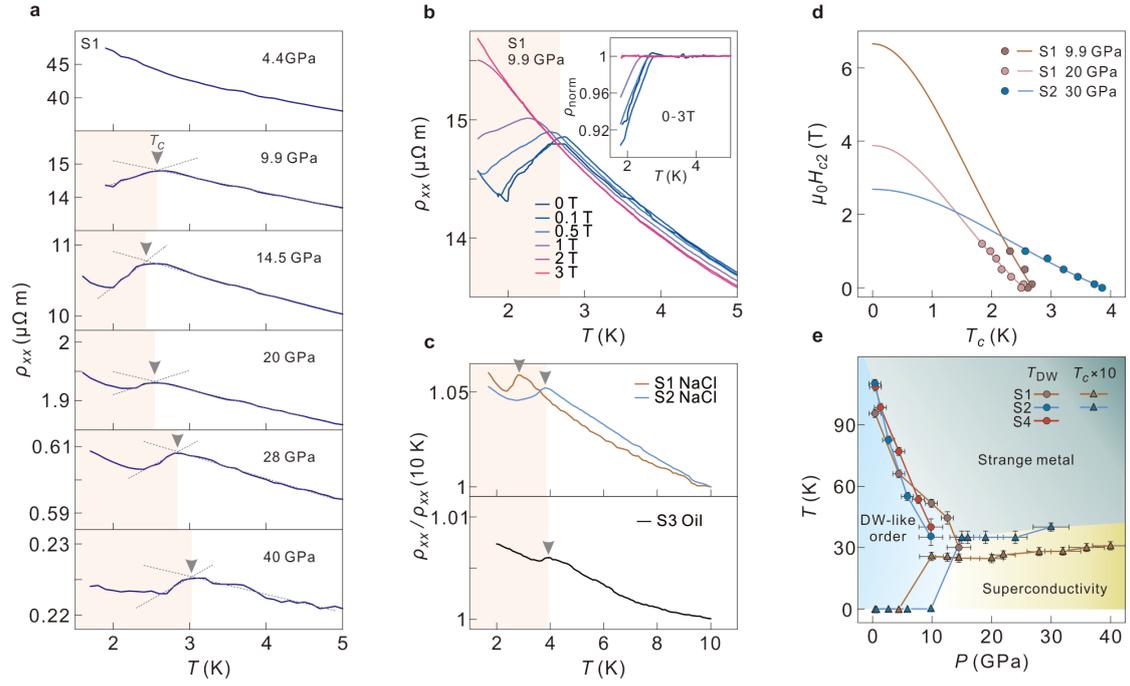

**Figure 4 | Emergence of superconducting-like transport under pressure. a**, Low-temperature $\rho_{xx}(T)$ of S1 at different pressures. **b,** $\rho_{xx}(T)$ of S2 at 9.9 GPa measured under magnetic fields from 0 to 3 T. The inset displays the normalized resistivity $\rho_{norm}$ from 0 to 3 T. **c,** $\rho_{xx}/\rho_{xx}(10\ K)$ for three samples measured under pressure with different pressure-transmitting media. S1 and S2 were loaded with NaCl, whereas S3 was loaded with silicone oil. **d,** Upper critical field $\mu_0 H_{c2}$ as a function of $T_c$ for S1 at 9.9 and 20 GPa, and S2 at 30 GPa. The symbols represent the experimentally determined characteristic fields, and the solid lines are fits using the three-dimensional Ginzburg–Landau formula. For S1 at 9.9 GPa, $T_c$=2.69±0.06 K，$\mu_0 H_{c2}(0)$=6.65±1.71 T；for S1 at 20 GPa, $T_c$=2.54±0.03 K，$\mu_0 H_{c2}(0)$=3.88±0.26 T；for S2 at 30 GPa, $T_c$=3.88±0.04 K，$\mu_0 H_{c2}(0)$=2.69±0.13 T. **e,** Pressure–temperature phase diagram of CVSO, showing suppression of the density-wave-like feature, the strange-metal-like regime and the emergence of the superconducting-like state under pressure. The characteristic temperatures, $T_{DW}$ and $T_c$, were determined from transport measurements on different samples. Lines are guides to the eye. Error bars indicate the estimated uncertainties in pressure and temperature.